\documentclass[12pt,reqno,a4paper]{amsart}
\usepackage{amsmath,amssymb,amsfonts}
\usepackage[mathscr]{eucal}

\newtheorem{theorem}{Theorem}

\theoremstyle{definition}
\newtheorem{remark}{Remark}
\newtheorem{example}{Example}[section]

\newcommand{\cal}{\EuScript}

\renewcommand{\L}{{\cal L}}

\newcommand{\p}{\partial}

\def\G {\Gamma}
\def\tnabla {\tilde{\nabla}}
\def\hnabla {\widehat {\nabla}}
\def\tG {\tilde{\Gamma}}
\def\hG {\widehat {\Gamma}}
\def\hM {{\widehat M}}

\def\X  {{\mathbf X}}
\def\hX  {{\widehat X}}
\def\Y  {{\bf Y}}

\def\F {{\cal F}}

\def\K {{\mathbf K}}
\def\hK {{\widehat K}}
\def\m{\medskip}

\def\l {{\lambda}}
\def\hw{{\widehat w}}

\def\hD{\widehat {\Delta}}
\def\wD #1 {{\widehat {\Delta_{#1}}}}

\def\hL {{\widehat {\cal L}}}

\newcommand{\bs}{{\boldsymbol{s}}}

\newcommand{\rh}{{\boldsymbol{\rho}}}

\title [] 
{Kaluza-Klein theory revisited: projective structures and 
differential operators on algebra of densities.}

\author{H.~M.~Khudaverdian}

\address{School of Mathematics,  University of Manchester,
 Oxford Road,  Manchester   M13 9PL,  UK}
\email{khudian@manchester.ac.uk }



\begin{document}

\maketitle

\begin{abstract}
We consider differential operators acting on densities of 
arbitrary weights on manifold $M$ identifying pencils of such operators
with operators on algebra of densities of all weights.
This algebra can be identified with the special subalgebra of functions
on extended manifold $\hM$.
On one hand there is a canonical lift of projective 
structures on $M$  to affine structures on extended manifold $\hM$. 
On the other hand the restriction of  algebra of 
all functions on extended manifold 
to this special subalgebra of functions 
implies the canonical scalar product. This leads in particular 
to classification of second order operators 
with use of Kaluza-Klein-like mechanisms.

\end{abstract}

\section {Algebra of densities}

In mathematical physics it is very useful to consider differential 
operators acting on densities of various weights on a manifold $M$
(see \cite{OvsTab} and citations there).
  We say that  $\bs=s(x)|Dx|^\l$ is a density of weight $\l$ on $M$
if under change of local coordinates  
     \begin{equation*}\label{transformationequations1}
                          \bs
           =s(x)|Dx|^\l=
s\left(x\left(x^\prime\right)\right)\left|\det
 \left({\p x\over \p x'}\right)\right|^\l
                    |D{x^\prime}|^\l\,,
        \end{equation*}
($\l$ is an arbitrary real number).
We denote by $\F_\l(M)$ the space of densities of weight
$\l$ on manifold $M$. (The space of functions on $M$ is $\F_0(M)$,
densities of weight $\l=0$.)

  Densities can be multiplied. If $\bs_1=s_1(x)|Dx|^{\l_1}$
and $\bs_2=s_2(x)|Dx|^{\l_2}$ are densities of weights $\l_1,\l_2$
respectively then  
   $
\bs=\bs_1\cdot\bs_2=s_1(x)s_2(x)|Dx|^{\l_1+\l_2}$
is a density of weight $\l_1+\l_2$.
We come to the algebra $\F(M)=\oplus_\l\F_\l(M)$ 
of finite linear combinations of densities of arbitrary
weights.
Use a formal variable $t$ instead volume form $|Dx|$.
Thus an arbitrary density 
$\bs=s_1(x)|Dx|^{\l_1}+\dots+s_k|Dx|^{\l_k}$ 
can be written as a function on $x,t$ which is 
{\it quasipolynomial} on $t$,
  $\bs(x,t)=s_1(x)t^{\l_1}+\dots+s_k(x)t^{\l_k}$.
An arbitrary density $\bs\in \F(M)$ can be identified 
with function
$\sum s_r(x)t^{\l_r}$ on the extended manifold $\hM$,
which is quasipolynomial on `vertical` variable $t$.
  There is a natural fibre bundle structure $\hM\to M$.
Extended manifold $\hM$ is the frame bundle of the 
determinant bundle  of $M$, $(x^i, t)$ are local coordinates on $\hM$.
Changing of local coordinates is:
     \begin{equation*}\label{transformationequations2}
(x^{i'},t')\colon\quad
   x^{i'}=x^{i'}(x^i),\quad t'=t'(x^{i},t)=
\det\left({\p x^{i'}\over \p x^{i}}\right)t\,.        
\end{equation*}

The fibre bundle $\hM\to M$ can be used for studying projective 
geometry on $M$ since there is a canonical construction which assigns to 
an arbitrary projective connection on manifold $M$ 
an usual affine connection on $\hM$ 
(see the last section).   
This affine connection on $\hM$ can be used for describing
the `projective geometry' on 
the base manifold $M$. Such investigation can
be traced to H.Weyl, Weblen and Thomas.
 On the other hand we will come to additional geometrical structures
on fibre bundle $\hM\to M$ if instead algebra of all smooth functions
on extended manifold $\hM$ we consider only the subalgebra of 
functions on $\hM$, which 
are quasipolynomial on vertical variable $t$, i.e.
  algebra $\F(M)$ of densities on $M$.
   This algebra can be equipped with  the canonical scalar
product:
If  $\bs_1=s_1(x)|D{x}|^{\l_1}$ and $\bs_2=s_2(x)|D{x}|^{\l_2}$ 
are two densities with a compact support then
                \begin{equation}\label{canonicalscalarproduct}
                   \langle \bs_1,\bs_2\rangle
                          =
                      \begin{cases}
                        & \int_M s_1(x)s_2(x)|D{x}|\,,
                 \quad {\rm if}\quad\l_1+\l_2=1\,,\cr
                                                               \cr
                        &  0\qquad 
          {\rm if}\qquad \l_1+\l_2\not=1\,.\cr
                      \end{cases}
                 \end{equation}
This construction  turns out to be very important tool to study geometry
of differential
operators on $M$ \cite{KhVor2}, \cite{KhVor4}. 
We  give short exposition of these results. Here we 
come to these results and formulate new ones 
based on alternative approach of Kaluza-Klein-like  reduction.  

\section {Differential operators on algebra of densities }

 Consider linear operator $\hw$ such that
$\hw(\bs)=\l \bs$
in the case if $\bs$ is a density of weight $\l$, ($\bs\in \F_\l(M)$).
If $\bs_1$ is a density of weight $\l_1$ and $\bs_2$ is a 
density of weight $\l_2$ then
          $$
  \hw(\bs_1\cdot\bs_2)=(\l_1+\l_2)\bs_1\cdot\bs_2=
  (\hw \bs_1)\cdot\bs_2+\bs_1\cdot(\hw \bs_2)\,.
          $$
Leibnitz rule is obeyed, $\hw$ is first order 
differential operator on the algebra of
densities. In local coordinates $(x^i,t)$ on $\hM$,
    $\hw=t{\p\over \p t}$.  A differential operator $\hD$
on algebra of densities has appearance 
$\hD=\hD\left(x,{\p\over \p x}, t, \hw\right)$ 
in local coordinates.
An arbitrary operator $\hD$ on algebra of densities defines the 
pencil of operators:
             $$ 
  \hD\mapsto  \{\Delta_\l\}\colon \Delta_\l=\hD\big\vert_{\hw=\l}\,.
                   $$
E.g. the operator  
$\hD=A(\hw)S^{ik}\p_i\p_k+B(\hw)T^{i}\p_i+C(\hw)R$
on algebra $\F(M)$
defines the pencil of operators
$\{\Delta_\l\}\colon$  
$\Delta_\l=A(\l)S^{ik}\p_i\p_k+B(\l)T^{i}\p_i+C(\l)R$.
Here $A,B,C$ are polynomials on $\hw$. If for example 
$A=1+\hw$, $B=\hw^2$ and $C=1$,
then $\hD$ is third order operator on the algebra of densities, which
defines the pencil of second order operators. 
Operators on algebra of densities can be identified with operator pencils
which depend polynomially on pencil parameter $\l$.

\begin{remark}\label{weight}
Here we consider only operators which do not change the weight
of densities: $\hD=\hD(x,{\p\over \p x},\hw)$, i.e. for corresponding
pencil $\{\Delta_\l\}$, $\Delta_\l\colon \F_\l(M)\to \F_\l(M)$.
\end{remark}

Canonical scalar product \eqref{canonicalscalarproduct} defines
adjointness of linear operators. 
Linear operator $\hD$ acting on the algebra
of densities has an adjoint $\hD^*$:
$\langle \hD \bs_1,\bs_2\rangle=\langle\bs_1,\hD^* \bs_2\rangle$.
  One can see that $(x^i)^*=x^i$, 
$\p_i^*=-\p_i$ and $\hw^*=1-\hw$.

   To consider self-adjoint and anti-self-adjoint operators
on extended manifold $\hM$ is very  
illuminating for studying  geometry of
operators on base manifold $M$. 
(See for details \cite{KhVor2}, \cite{KhVor4},\cite{KhBiggs1}
and \cite{KhVor5}.)

\section {First order operators and Kaluza-Klein mechanism}

Consider an arbitrary 
first order operator $\widehat K$ such that it does not change the
weight of densities (see Remark \ref{weight}) and obeys normalisation
condition
$\hK(1)=0$. In local coordinates it has the following appearance
             \begin{equation*}
       \hK=K^i(x)\p_i+ K^0(x)\hw\,.
           \end{equation*} 
One can see that its adjoint is equal to
$\hK^*=-K^i(x)\p_i-\p_iK^i(x)+ K^0(x)(1-\hw)$. 
$\hK$ is a vector field on extended manifold $\hM$. 
One can define divergence
of this vector field:
             \begin{equation}\label{divergence}
       {\rm div\,}\hK=-(\K+\K^*)=
   \p_i K^i(x)- K^0(x)\,.
           \end{equation}    
Notice that the divergence is defined in spite of 
the absence of well-defined volume form on $\hM$ 
(see for details \cite{KhVor2} and \cite{KhVor5}).

  We see that 
vector field $\hK$ is divergenceless iff it is anti-self-adjoint:
            \begin{equation*}\label{liederivative}
\hK=-\hK^*\Leftrightarrow {\rm div\,}\hK=0
\Leftrightarrow     \hK=K^i(x)\p_i+\hw\p_i K^i(x)\,. 
          \end{equation*}  
One can see that divergenceless 
vector field $\hK$ is a Lie derivative
of densities along its projection, vector field $\K$ on $M$. 
An arbitrary vector field
$\X=X^i(x)\p_i$ can be lifted to anti-self-adjoint 
(i.e. divergenceless) vector field on extended manifold $\hM$, 
which is nothing but Lie
derivative of densities: $\X\mapsto \hL_\X$ such that
for arbitrary $\bs=s(x)|Dx|^\l$,
            \begin{equation*}\label{liederivative2}
    \hL_\X(\bs)=\widehat X(\bs)=
\left(X^i(x)\p_i+\hw\p_iX^i(x)\right)s(x)|Dx|^\l=
\left(X^i\p_is(x)+\l\p_i X^is(x)\right)|Dx|^\l\,.
          \end{equation*}  
It is useful to consider a connection on a bundle $\hM\to M$.
 It assigns to every vector field $\X=X^i(x)\p_i$ 
on $M$ its lifting, the horizontal vector field 
$\hX_{\rm hor}=X^i(x)\p_i+\gamma_i(x)X^i(x)\hw$ on $\hM$.
Connection defines derivation $\nabla_\X$ on algebra of densities:
for $\bs=s(x)|Dx|^\l$
         \begin{equation}\label{connection1}
  \nabla_\X (\bs )=\hX_{\rm hor}(\bs )=
         \left(       
X^i(x)\p_i s(x)+\l\gamma_i(x)X^i(x)s(x)
          \right)|Dx|^\l\,.
     \end{equation}
Under changing of local coordinates $x^i\to x^{i'}=x^{i'}(x^i)$
components
$\gamma_i$ of connection are transforming in the following way:
           \begin{equation*}
\gamma_{i'}={\p x^i\over \p x^{i'}}
                 \left(
         \gamma_i+{\p\over \p x^{i}}
                 \log\left(\det{\p x^{j'}\over \p x^j}
                 \right)
                 \right)\,,\quad
      \left(\gamma_i(x)|Dx|=\nabla_i|Dx|\right)\,.
\end{equation*}
Connection $\gamma_i(x)$ defines divergence ${\rm div\,}_\gamma\,$ 
of vector fields on $M$, 
which is equal to divergence \eqref{divergence} 
of horizontal lifting of this vector field:
            ${\rm div\,}_\gamma \X={\rm div\,}\hX_{hor}=
\p_i X^i(x)-\gamma_i(x)X^i(x)$.   

\begin{remark}
Let $\X=X^i\p_i$ be a projection on $M$ of
a vector field $\hX=X^i\p_i+\hw X^0$ on $M$, 
and $\hX_{\rm hor}$ be a horizontal lifting of $\X$. Then
$\hX-\hX_{\rm hor}=\hw(X^0-\gamma_iX^i)$ is a vertical vector field
and $X^0-\gamma_iX^i$ is a scalar field.
\end{remark}

\begin {remark}\label{christoffel}
A volume form  $\rh=\rho(x)|Dx|$ on $M$ naturally defines
a connection $\gamma_i=-\p_i \log\rho(x)$.
A Riemannian metric $G=g_{ik}dx^idx^k$ on $M$ 
naturally defines a volume form $\rh=\sqrt{\det g}|Dx|$.
The corresponding connection $\gamma_i=-\Gamma^k_{ik}$,
where  $\Gamma^i_{km}$ are
Cristoffel symbols of Levi-Civita connection of the metric.
In this case ${\rm div\,}_\gamma\,$ is a standard divergence operator
(with respect to a volume form).
\end{remark}

\section { Second  order operators and Kaluza-Klein reduction}

  Let $\hD$ be an arbitrary second order operator on algebra of densities
$\F(M)$:
            \begin{equation}\label{secondorderoperator1}
   \hD=
    \underbrace{S^{ik}(x)\p_i\p_k+
   2 \hw B^i(x)\p_i+\hw^2 C(x)}_{\hbox{second order derivatives}}+
\underbrace{D^i(x)\p_i+\hat w E(x)}_{\hbox{first order derivatives}}+
          F(x)\,.
             \end{equation}
(As always we consider only
operators which do not change weight of densities (see remark \ref{weight}).)
  
  Principal symbol of this  operator is 
          \begin{equation*}\label{principalsymbol}
       \widehat S=\begin{pmatrix}
               S^{ik}& B^i\cr
                   B^k& C  \cr
          \end{pmatrix}\,,\qquad
    (\hbox{in coordinates\,\,} x^i,x^0=\log t)\,.
         \end{equation*}
  $\widehat S$ is a contravariant symmetric tensor field on the
extended manifold $\hM$. 
`Space components' $S^{ik}$ of the tensor field $\widehat S$
are components of symmetric contravariant tensor field on $M$.
Operator $\hD$ defines a pencil of second
order operators $\{\Delta_\l\}$, $\Delta_\l=S^{ik}\p_i\p_k+\dots$,
and all these operators have the same principal symbol $S^{ik}$.

Put normalisation condition $F=\hD(1)=0$ and consider the operator
which is adjoint to operator \eqref{secondorderoperator1}: 
                  $$
             \hD^*=
               \p_k\p_i
            \left(S^{ik}\dots\right)-
     \hw^* \p_i
          \left(2B^i + (\dots)\right)+
            (\hw^*)^2 \left(C \dots\right)-
            \p_i\left(D^i \dots\right)+
      \hw^* E \,,\,(\hw^*=1-\hw)\,.
        $$
The condition that operator $\hD$ is self-adjoint, $\hD^*=\hD$ 
implies that
              \begin{equation}\label{selfadjointoperator}
  \hD=  
    S^{ik}\p_i\p_k+\p_k S^{ki}\p_i+
  \left(2\hw-1\right) B^i\p_i+\hat w\p_i B^i+
         \hw \left(\hw-1\right)C\,.
                \end{equation}
Thus self-adjoint second order operator on algebra of densities,
which obeys normalisation condition
$\hD(1)=0$ is uniquely defined by its symbol.   

The geometry of operator \eqref{selfadjointoperator} was studied in detail
in articles \cite{KhVor2}, \cite{KhVor4} and \cite{KhBiggs1}.
  Here we present   and analyze 
  these results, 
using Kaluza-Klein-like  mechanism.

  Kaluza-Klein mechanism defines a connection (gauge field) 
and Riemannian metric
on a base manifold  through Riemannian metric on 
a total space of fibre bundle $\hM\to M$.
Connection, i.e. the distribution of horizontal hyperplanes
(subspaces 
 which are transversal to the fibres) is 
defined by the condition that these hyperplanes are orthogonal to the fibres
with respect to Riemannian metric in the bundle space.

One can slightly alter this mechanism. 
  Contravariant tensor field $\widehat S$, principal symbol of operator 
\eqref{selfadjointoperator} maps $1$-forms (covectors) 
to vectors on $\hM$.   Consider the following 
Kaluza-Klein-like mechanism: take an arbitrary $1$-form $\Omega$ on 
$\hM$ such that $\Omega(\hw)\not=0$, i.e.
$\Omega$ is proportional to form $d x^0+\dots$ 
($x^0=\log t$), and the following condition is obeyed:
vector field $\widehat S \Omega$ is proportional to vertical vector
field $\hw$ ($\hw=t{\p\over \p t}={\p\over \p x^0})$ .
 This means that for $1$-form $\Omega=a(x)(dx^0-\gamma_k(x)dx^k)$
the following condition holds:
  \begin{equation}\label{connectionform}
               \begin{pmatrix}
               S^{ik}& B^i\cr
                   B^k& C  \cr
               \end{pmatrix}
               \begin{pmatrix}
              -\gamma_k\cr
                   1\cr
               \end{pmatrix} 
                 \,\hbox {is proportional to vector}\,
               \begin{pmatrix}
               0\cr
                1\cr
               \end{pmatrix} 
                   \,. 
            \end{equation}
This condition canonically 
defines distribution of horizontal hyperplanes in $\hM$,
which are sets of vectors which annihilate the form $\Omega$.
 (One can take $\widehat S\Omega=0$ 
in the case if $\widehat S$ is degenerate.) 
 Every vector field $\X=X^i(x)(x)\p_i$ on the base $M$ can be
lifted to horizontal vector field which annihilates the connection form 
$\Omega$ (see also equation \eqref{connection1}). 
This construction works if condition \eqref{connectionform} 
is obeyed, i.e. in the case if the equation 
         \begin{equation}\label{equationshavesolution}
          S^{ik}(x)\gamma_k(x)=B^i(x)
          \end{equation}
has a solution. In this case  
second order operator \eqref{secondorderoperator1}
via its principal symbol $\widehat S$
defines a connection $\gamma_k$. In the case if $S^{ik}$
is non-degenerate then an equation \eqref{equationshavesolution}
has unique solution. In this case operator defines 
uniquely canonical connection and Riemannian metric on the base.

 The field 
$B^i(x)=S^{ik}(x)\gamma_k(x)$ can be considered
as an {\it upper connection}. 
It follows from \eqref{connectionform}
 that 
in this case $C-B^i\gamma_i$ is a scalar.
and $C$ is related with
{\it Brans-Dicke function}.

    In general case if the condition \eqref{equationshavesolution}
is not obeyed then more detailed analysis shows
that $B^i-S^{ik}\Gamma_k$ is a vector field and 
$C-2B^i\Gamma_i+S^{ik}\Gamma_i\Gamma_k$ is a scalar,
where $\Gamma_i$ is an arbitrary connection.

The importance of operator \eqref{selfadjointoperator} 
is defined by the following 
uniqueness Theorem:

\begin{theorem} 
Let $\Delta$ be second order operator acting on 
densities of weight $\l_0$,
where $\l_0\not=0,1,1/2$.

Then there exists a unique self-adjoint operator  $\hD$
($\hD^*=\hD$) which obeys the following conditions
   \begin{itemize}

\item
             $
           \hD\big\vert_{\hat w=\l_0}=\Delta
             $,

\item  normalisation condition $\hD(1)=0$. 

\end{itemize}

In other words there exists unique self-adjoint 
normalised pencil of second
order operators which passes through a given operator.

\end{theorem}

This Theorem was formulated and proved in \cite{KhVor2} (see also
 \cite{KhVor4}).

\begin{example}  Consider second order operator
      $
\Delta_{\l_0}=\L_\X\circ\L_\Y
      $,
where $\X,\Y$ are arbitrary vector fields,
$\L_\X,\L_\Y$ are Lie derivatives of densities of weight $\l_0$
along vector fields $\X,\Y$ respectively. 
  Construct the following  operator on algebra of densities:
         \begin{equation*}
  \hD=
    {1\over 2}\left(
   {\widehat \L_\X}{\widehat \L_\Y}+
   {\widehat \L_\Y}{\widehat \L_\X}
       \right)+
         {1\over 2}
       \left({2\hw-1\over 2\l_0-1}\right)
       \left(
   {\widehat \L_\X}{\widehat \L_\Y}-
   {\widehat \L_\Y}{\widehat \L_\X}
       \right)\,.
         \end{equation*}
This operator is obviously self-adjoint operator and it passes through
operator $\Delta_{\l_0}$. One can see that it is equal to 
          \begin{equation*}
                \hD=
   {\widehat \L_\X}{\widehat \L_\Y}+
       \left({\hw-\l_0\over 2\l_0-1}\right)
          \widehat \L_{[\X,\Y]}\,.
         \end{equation*}
         \end{example}

Let $\Delta_{\l_0}=A^{ij}\p_i\p_j+A^i\p_i+A$ 
be an arbitrary second order operator
acting on space of densities of a given weight $\l_0$, ($\l_0\not=0,1/2,1$). Then we see that the self-adjoint operator \eqref{selfadjointoperator} 
passes via the operator $\Delta_{\l_0}$ if for upper connection $B^i$
and Brans-Dicke function $C$ 
           $$
   B^i={A^i-\p_kA^{ki}\over 2\l_0-1}\,, 
C={A\over \l_0(\l_0-1)}-
  {\p_iA^i-\p_i\p_kA^{ki}\over (\l_0-1)(2\l_0-1)}\,.
           $$
The upper connection $B^i$ 
is induced by a connection $\gamma_i$ ($B^i=A^{ik}\gamma_k$) iff  
equation \eqref{equationshavesolution} has a solution 
(for $S^{ik}=A^{ik})$.
 The condition \eqref{equationshavesolution} defines this special
property of second order operators on densities. 
It is interesting to analyze its
geometrical meaning.

\section {Thomas bundle and projective geometry}

   The canonical constructions  studied  in the previous sections were
  successfully performed since we consider not 
the algebra of all functions
on extended manifold  $\hM$, but only functions which are 
quasipolynomial on vertical variable $t$, since 
scalar product \eqref{canonicalscalarproduct}
is not well-defined on algebra of all (smooth) functions
on $x,t$. Nevertheless in  general case
for fibre bundle $\hM\to M$
there exists the remarkable construction which assigns 
to projective class of connections on $M$ 
the affine connection on $\hM$.   
This construction is due to T.Y.Thomas \cite{Th1}.
(See also \cite{J1} and \cite{J2}).
The bundle $\hM\to M$ sometimes is called Thomas bundle.
Now we sketch this construction.

   We say that two symmetric affine connections
$\nabla$ and $\tnabla$ on manifold $M$ 
belong to the same projective class  $[\nabla]=[\tnabla]$ if
         $$
      \tnabla-\nabla
 =\tG^i_{km}-\G^i_{km}=t_k\delta^i_m+t_m\delta^i_k\,,\qquad
(t_i \,\hbox{is covector})\,,
           $$
where 
$\tG^i_{km}$ and $\G^i_{km}$ are Christoffel symbols of connections
$\tnabla$ and $\nabla$ respectively. 
  Equivalence class of symmetric connections is projective connection.
(Projective connection in particular defines non-parametrised geodesics:
two symmetric affine connection belong to the same class iff 
they have the same non-parameterised geodesics.)
 
  For affine connection $\nabla$ on $n$-dimensional
 manifold $M$ 
with Christoffel symbols
  $\G^i_{km}$ one can consider symbols
    \begin{equation}\label{projconnection2}
  \Pi^i_{km}(\nabla)=\Pi^i_{km}=\G^i_{km}+
 {1\over n+1}\left(\gamma_k\delta^i_m+\gamma_m\delta^i_k\right)\,,
    \end{equation} 
where $\gamma_i=-\G^k_{ik}$ define connection on densities on $M$
 (see also remark \ref{christoffel}).
   Two symmetric
connections $\nabla,\tnabla$ belong to the same projective class
iff  $\Pi^i_{km}(\nabla)=\Pi^i_{km}(\tnabla)$.

  Let $[\nabla]$ be a projective class of symmetric connections
  on $n$-dimensional manifold $M$.
Then Thomas construction assigns to this projective class $[\nabla]$
 the symmetric affine
connection $\hnabla$ on the extended manifold $\hM$.   
with following Christoffel symbols 
     $$
\hG^i_{km}=\Pi^i_{km},\, 
   \hG^0_{km}={1\over n+1}\left(\p_r\Pi^r_{km}-\Pi^r_{sk}\Pi^s_{rm}\right)\,,
\hG^i_{k0}=-{\delta^i_k\over n+1}, 
   \hG^i_{00}=\hG^0_{i0}=0,
    \hG^0_{00}=-{1\over n+1}\,.
         $$
	 Here $\Pi^i_{km}$ are symbols 
\eqref{projconnection2} corresponding
to Christoffel symbols of a connection in the class $[\nabla]$.
(We use local coordinates $(x^i,x^0)=(x^i,\log t)$) 
in the extended space.)

\m
  
\textbf{Acknowledgments}
This article is based on my talk at the conference
``The Modern Physics of Compact Stars and Relativistic Gravity''
 (Yerevan, Septemebr 2013).
I am grateful to organisers of this conference
for financial support,  and also to Engineering and Physical
Sciences Research Council (EPSRC) for financial support
via the Mathematics Platform Grant (MAPLE) EP/IO1912X/1.
I am grateful to T.Voronov for fruitful discussions.

\end{document}